# *27*

# Big Data: How Geo-information Helped Shape the Future of Data Engineering

Robert Jeansoulin

CNRS senior researcher, Université Paris-Est Marne-la-Vallée

Laboratoire d'informatique, Institut Gaspard Monge

(On leave, as: Science Advisor at French General Consulate in Quebec)

robert.jeansoulin@univ-mlv.fr

**ABSTRACT.** Very large data sets are the common rule in automated mapping, GIS, remote sensing, and what we can name geo-information. Indeed, in 1983 Landsat was already delivering gigabytes of data, and other sensors were in orbit or ready for launch, and a tantamount of cartographic data was being digitized. The retrospective paper revisits several issues that geo-information sciences had to face from the early stages on, including: structure  ( to bring some structure to the data registered from a sampled signal, metadata); processing  (huge amounts of data for big computers and fast algorithms); uncertainty (the kinds of errors, their quantification); consistency (when merging different sources of data is logically allowed, and meaningful); ontologies (clear and agreed shared definitions, if any kind of decision should be based upon them). All these issues are the background of Internet queries, and the underlying technology has been shaped during those years when geo-information engineering emerged.

**A. Title of AutoCarto Six paper.**

Automatic cartography of agricultural zones by means of multi-temporal segmentation of remote sensing images.

**B. Reason for paper?**

In 1983, working on the above title, the purpose of my then sponsor[1] was to prepare the launch of satellite SPOT and the forthcoming commercialization of its products, with a focus on vegetation monitoring.

It may look unbelievable today, but we were not equipped with image-capable screens, only alphanumeric consoles: everything had to be printed for being displayed. However, data were there, big matrices of data, that couldn't be turned easily into images.

---

[1] Centre National d'Etudes Spatiales, Département de Traitement de l'image, Toulouse, France.

Therefore, we were forced to crunch data, failing to be able to look at them! The amount of images that a satellite such as Landsat was able to harvest was phenomenal: gigabytes, terabytes of pixels: we were using the super computers of that time.

Dealing with "big data" before the term was popularized? The armory of mathematical tools also was almost there: principal component analysis, multi-dimensional correlation, template matching, and so on. We may say that in remote sensing, in photogrammetric engineering, in geographical information systems, or for short in geo-information (or geomatics), we pioneered what today is termed "big data".

This paper browses the principal issues that the geo-information science had to face from the early stages on. First challenge: to bring structure to the data registered from a sampled signal, what eventually gave metadata and the ability to merge images into large retrieval systems in the Internet. Other challenges involved: processing (such huge amounts of data required big computers and fast algorithms); data uncertainty (the description of the kind of errors, and their quantification was necessary from the very beginning); data consistency (as soon as we started merging different sources of data, it became mandatory to question if and how far we were allowed to merge them. In French we say, M*ariage de la carpe et du lapin*, (Carp and rabbit wedding).

Finally, ontology questions are addressed, because the comparison of geo-data, piled up for several decades, all around the globe, imposes clearer definition on what they are data of or what they represent, what comparison can be made, what real evolution or differences they measure, and what kind of decision can we base upon them.

Today, these same longstanding issues and questions continue to be raised in the context of big data.

**C. Data structure: From ancillary data to metadata.**

Remote sensing imagery inaugurated the use of metadata as soon as the first images had to be registered, overlaid, and when different sensors were used.

Basically we have only two kinds of data: words and numbers. And if we have a number, we do need words too: if 24, 24 what? Natural or physical sciences deal with numbers, with signal or image processing, with geometry, time series, etc. But text processing is the main approach in big data: "*In principio erat Verbum*", the very first words of the Gospel of John. The two didn't fit well 30 years ago, but nowadays the difference is not so drastic.

Noticeably, today, is the widespread use of metadata. I do remember that we didn't use the word metadata but "ancillary data", to name the data associated with remote sensing imagery: for instance the ground resolution, wavelengths, etc. (see below). The name change denotes an upgrade for something secondary (*ancillary*) to a more important role (*meta*). The way we consider data has evolved as well.

For instance: pixel data. A pixel is the approximation of a ground surface, from which we measure a reflected (or emitted) radiometry within a certain wavelength range, integrating diffraction effect, absorption, etc. As many of us did, I spent a lot of time modeling these effects, filtering them to improve the data from a status of "raw data" to a status of "corrected data". Also, pixels aren't processed one by one, but as a statistical variable that receives class membership or geometrical properties (e.g. border pixel). These properties must be described and registered into some structure: a "processed image" has a lot of such attached information (metadata).

Libraries were confronted by the metadata issue and developed MARC in the sixties, a markup language to which HTML owes a lot. An important next step was the Dublin Core (DCMI) in 1995. In automated cartography, one big question was, How to introduce the topology in the data vector representation? It's implicit from the geometry, but the burden of re-computing it is much too heavy. Then, in the 1990s several NGOs[2] were working on what became ISO 19101: 2002 Geographic Information Reference model.

For instance, Table 1 represents the geometry and the topology of a set of land parcels and allows determining that the union of #2 and #3 forms a single hole into #1.

Table 1. A relational representation of the (polygon-arc-node) topology schema.

| # | Coordinates (or vertices) | contains | is in | touches | has hole | … more … |
|---|---|---|---|---|---|---|
| 1 | *x,y; x,y; x,y ; x,y; x,y …* | 2;3 | - | - | 1 | … |
| 2 | *x,y; x,y; x,y; x,y …* | - | 1 | 3 | 0 | … |
| 3 | *x,y; x,y; x,y …* | - | 1 | 2 | 0 | … |

The content of such tables is described in the ISO reference model: the polygon-arc-node model, with all topology relationships. Moreover, semantics and rules can be added too (See F, Data consistency.).

In big data, there are trillions of sparse, scattered, unrelated docs (unstructured) anywhere on the Internet, and the goal of so-called "robots" is to attach structure (the indexing process) and try to rank them in response to a single request (querying the Internet), or to a complex and multi-morphed request (data analytics).

The concept of unstructured data wasn't in use 30 years ago. Relational databases were just on the rise (and still are embedded in most data engineering). The term *semi-structured data* appeared in 1995 in the community of database specialists, and XML was first established in 1997 (built upon the SGML of the late 1980s).

Hence, there are reasons to consider that data processing of signal and image is one of the many precursors of today's big data.

**D. Data processing: From data analysis to data mining.**

Signal processing has been a field of experimentation of innovative statistical, stochastic, or data transformation approaches, as well as intensive computational methods and efficient algorithms for the processing of very large data sets. Exploratory data analysis (EDA) is linked with Tukey and Fourier analysis and similar approaches, developed mainly in the field of signal processing. The term is outdated now, but it certainly contributed to foster research in high-end computing. Let's recall for instance the hierarchical classification algorithms used for pixel classification in Landsat images (Jeansoulin 1981), and the use of more sophisticated approaches for unsupervised classification, such as the "dynamic clusters" (Diday 1973). Figure 1 illustrates the basics of the process.

---

[2] National geographic offices, such as Ordnance Survey, IGN Geomatics Canada, USGS, etc.

Figure 1. Hierarchical clustering example (sometimes referred to as phylogenetic tree).

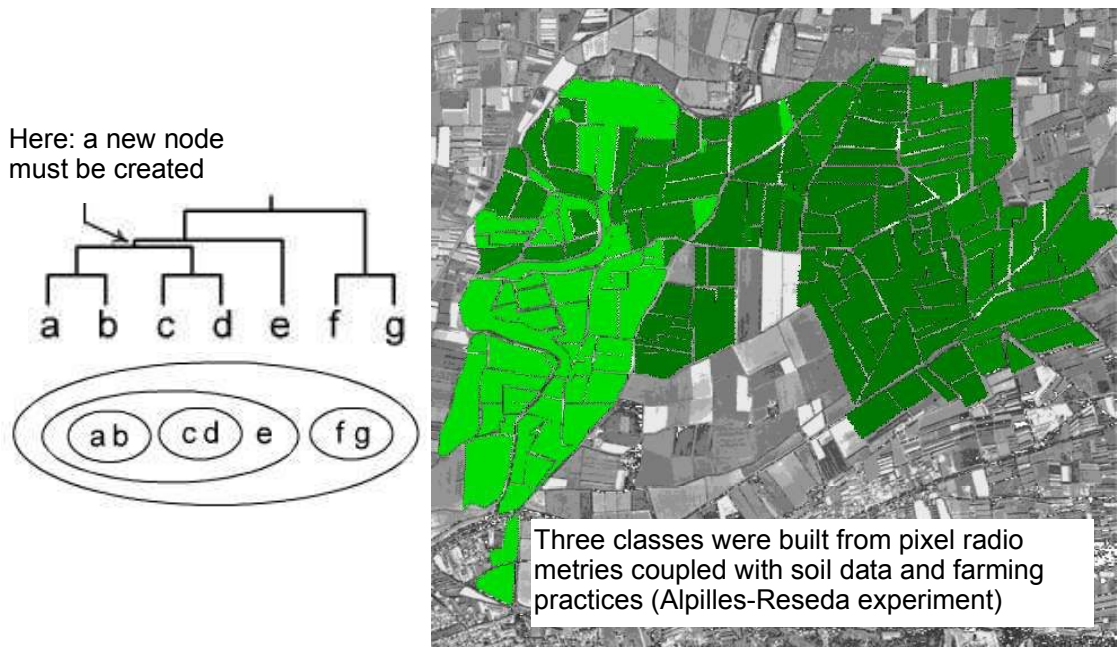

Three classes were built from pixel radio metries coupled with soil data and farming practices (Alpilles-Reseda experiment)

The descending operation tries to split the data domain into relevant sub-classes at each level, leading to a tree structure (See section G, Ontologies.). It is then applied to field parcels according to their median or average pixel value (Olioso 1998).

The expression *data mining* traces back to the 1990s, familiar to database scientists, and mainly as an operational approach of machine learning whose foundations are linked to the Turing machine: the field was more theoretical, between logics and computing languages, lambda-calculus. Support vector machine (SVM) algorithms were developed as non-probabilistic binary linear classifier (Vapnik 1995).

Nowadays, these terms have been more or less wrapped up in the successive buzzwords of business intelligence, data analytics, and big data. In spite of other differences, the fact is that the mainstream shifted from signal processing to the world of Internet and e-commerce.

But, looking deep into the algorithms it can be seen that: the geo-processing legacy is there. From a computational point of view (supercomputer and huge data storage), or with respect to the underlying exploratory approaches, geo-processing research has contributed to paving the way.

**E. Data uncertainty: From precision to quality indicators.**

Geo-information deals with the "real world": without arguing about philosophical truth, it deals at least with a same and single "world" which can be modeled and measured by different means, at different scales, from multiple viewpoints, etc., and everything has to be logically consistent.

Therefore, geo-processing talks not only about a particular aspect of reality, but also about a consistent framework for this reality.

At first, measurements quality was focused on precision: What particular location on Earth does a particular pixel represent? What is the real radiometric contribution of this location actually represented in the pixel value? Image registration, and sensor fusion, were the first and most difficult tasks at the time. Soon after that, confidence involving data classification was the big issue, e.g., *How much wheat is the USSR really harvesting*?

Questions of that nature were investigated by relatively the same amount of computing power that the NSA can harness today for processing trillions of emails.

When it became easier to merge remote sensing imagery and geographical databases, much more complex questions were at hand. The gain in ground resolution enabled us to reach new terrain: from the large rectangular crops of the Mid-West to small vineyards in Provence, and now to urban roof gardens. Sociology is no longer some weird science taught to hippies, but a science we can have trans-disciplinary talks with, as we did with agronomy.

The challenge about quality is no solely longer about data precision (though still important), but also about consistency: Are we talking about the same thing when merging different data?

Geo-processing has extensively investigated the question of quality, with the result that a number of quality indicators have been designed, and the quality domain has been structured by international consensus.

Since the 1990s, geo-information specialists met in consortia such as OGC, and eventually established an ISO Technical Committee (ISO TC211) to discuss and publish standards for geo-information (geomatics): data models and data quality information were among the most specific outcomes: ISO 19101: "reference model", and ISO 19113: "quality principles" were finally issued in 2002, reflecting common ground between the various national geographic organizations (see Table 2).

The national statistics agencies were closely working with these specialists, because the same quality issues are important for the countries and for international comparison as well. International bodies such as UNESCO, OECD, and Eurostat have also been aware of the same issue for many years, but the automation of cartography was probably among the very first pioneers.

Table 2. The actual consensus about metadata for quality in ISO19113:2002.

| Data quality element / sub-element | Description |
|---|---|
| Completeness (omission, commission, logical consistency) | Presence of features, their attributes and relationships (absence or excess of data, adherence to rules of the data structure) |
| Conceptual consistency | Adherence to rules of the conceptual schema, to value domains, etc. |
| Topological consistency | Correctness of explicit topology, closeness to respective position of features |
| Positional accuracy | Accuracy in absolute point positioning, gridded data positioning |
| Temporal consistency | Accuracy of temporal attributes and their relationships |
| Thematic accuracy | Accuracy of quantitative attributes, class correctness |

**F. Data consistency: Uncertain but rational knowledge.**

Understanding the many causes of data uncertainty sheds light on the many approximations made all along the process of gathering and measuring even the simplest datum: for instance, ground temperature.

Considering that data are always somewhat inexact, and considering that data always depend on a model imperfectly representing a reduced aspect of reality, it is important to provide guidelines or constraints. Every time we can provide some constraints we can confront the data, and issue a warning for each detected conflict.

Figure 2 is a snapshot of a successful experiment developed during the FP5 European project REVIG!S[3]: how to revise – by logical means and tools – very uncertain flood data using direction of water flow as constraints (Jeansoulin and Wilson 2002).

In big data, the prediction of the annual flu wave by looking through Internet "medical queries" has got a lot of media attention.

It isn't necessarily a convincing example, but the archetypical "analytics" story is IBM Watson, when it overcomes two Jeopardy champions, in 2011. The DeepQA project behind Watson is making intense use of geo-information reasoning for answering questions such as "*They're the two states you could be reentering if you're crossing Florida's northern border*" (Ferrucci 2010).

Mereotopology, region connectedness calculus (RCC) or Allen's interval algebra [Allen 1983], developed in the 1980s, are extensively used today for constraining queries and making answers more narrow and efficient.

---

[3] REVIGIS project involved universities of Marseilles, Leicester, Keele, Technical Vienna, Pisa, Twente (ITC), Laval.

Figure 2. Integrating data and constraints: flow direction information and estimated water heights must comply or be "revised".

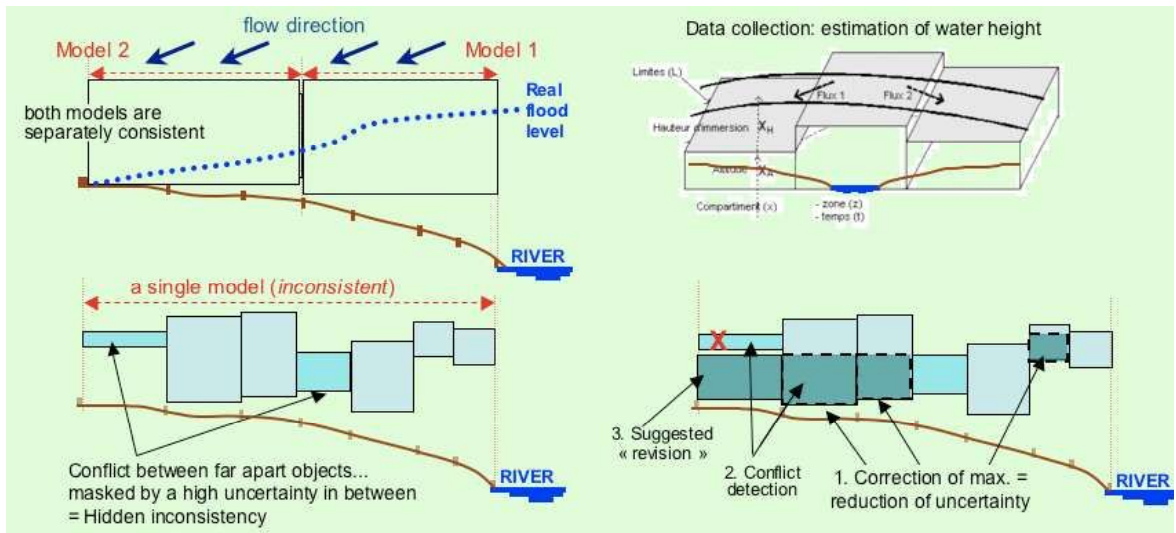

Figure 3 gives the lattice the 8 possible such algebras (Euzenat 1997).

Figure 3. Lattice of 8 mereotopology algebras.

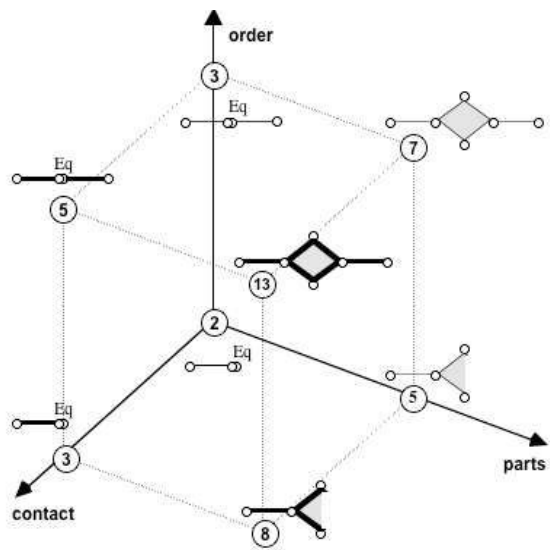

Description. Each node gives the number of allowed relations per algebra (from 2 to 13); and the axis have the following effects:

*contact* introduces arc (bold) relations,

*parts* explodes the equivalence relation Eq,

*order* unwraps the graph around a symmetry axis (Allen 1983).

Reasoning under constraints marries well with stochastic reasoning, as with Bayes networks (Cavarroc 2004). In Figure 4, the network (1) is the direct application of the Pearl algorithm on data (pixel data of two images 1995 and 1996, plus terrain data (altitude, pedology, etc.); the constrained network (2) accepts some additional constraints in the algorithm (e.g. slope doesn't impact the upstream/downstream rank in the valley, or landcover is not a cause for area); the constrained network is used to build a prediction (3) for the following year, from the land use in 1995 (separate data set).

Figure 4. Bayes networks used in prediction mode.

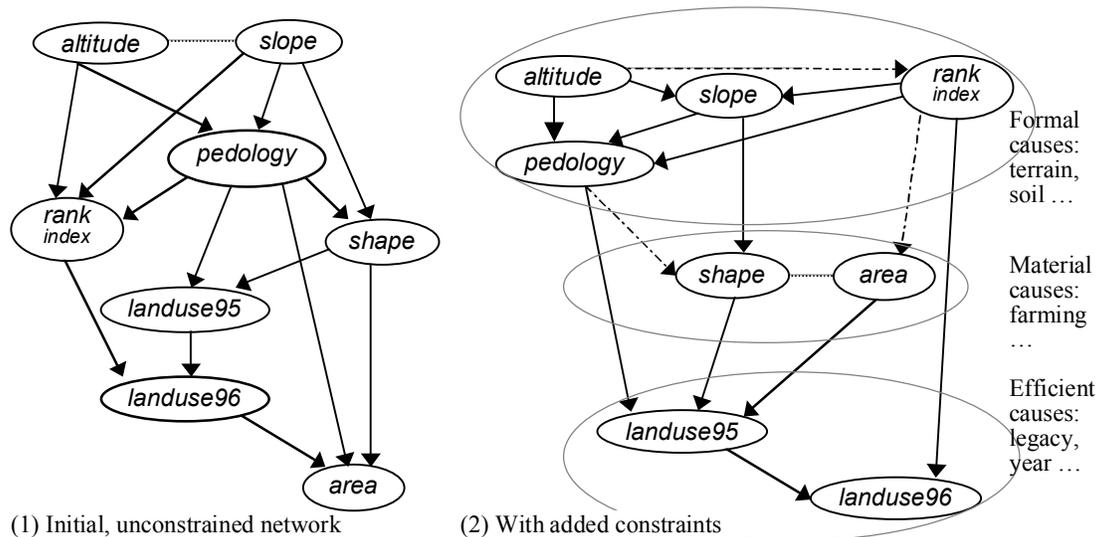

(1) Initial, unconstrained network    (2) With added constraints

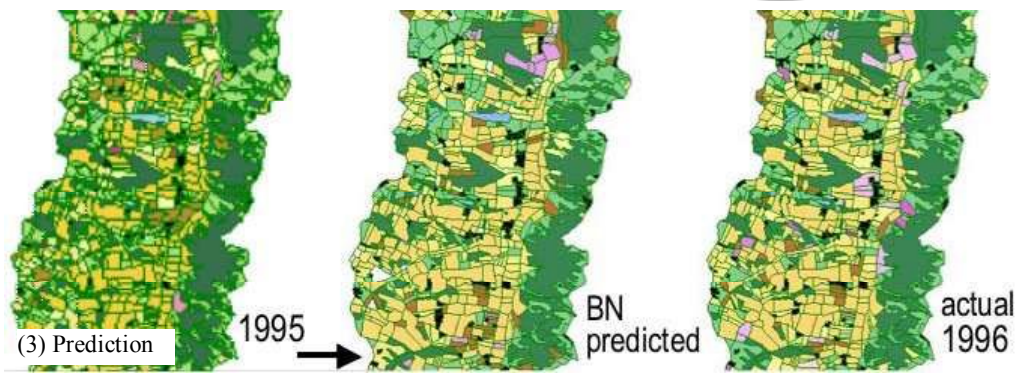

(3) Prediction

## G. Ontologies: Data are acts, not facts.

Geo scientists are still classifying and zoning, but much more attention is turned to the meaning of the process, the interpretability of the result, and the ability to use it within a decision process. Moreover, geo-information also raised the question of what is in data, common agreement, external quality, data usability, etc., which are different aspects of a more general question often summarized into the word "ontologies" (plural!), and the subsequent problem of "ontology alignment" (Gruber 1994; Halevy 2005).

Here again, the terminology has been brought to public attention by the wide spread of the Internet and the mainstream studies following it, but, here again, geo-information was developing its research when questioning the problems of global data quality, of managing different quality levels for different types of information, and geomatics was among the first to provide a systematic and comprehensive approach, with the above-mentioned ISO 191xx series of standards, in particular in 2002 with ISO 19150 ("ontologies"), and ISO 19157 ("data quality"). Deviller and Jeansoulin (2006) regroup a good part of this research on ontologies and quality, in particular the notion of external quality (quality for the user) which is illustrated in Figure 5, where the quality defined by the producer of geo-data is internal (referring to its own specification) rather than external (quality as a fitness for use).

Figure 5. The internal versus external quality issue.

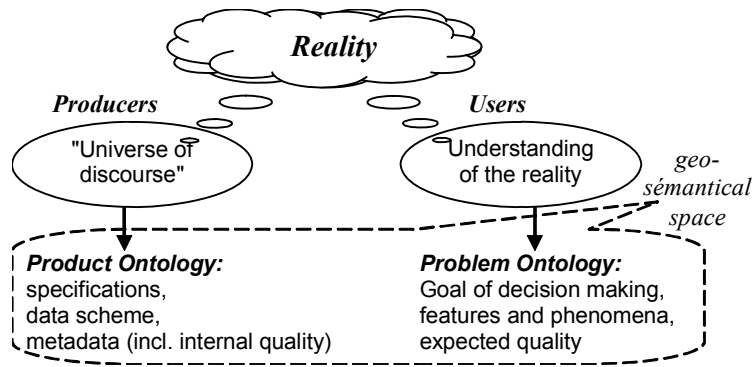

Let's develop an example of ontologies alignment with agricultural data. Given 2 graphs representing two different surveys of the same region (Pham 2004), the observations at the same locations (parcels) will differ according to their ontologies, and we can build a Galois lattice summing up the fusion of information.

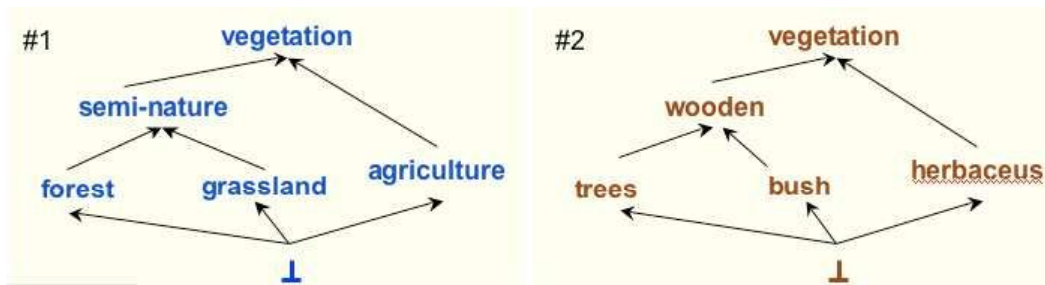

In Figure 6, the conflated graph contains all the different nodes of both initial graphs, plus some additional mandatory nodes (yellow), minimal addition for a consistent fusion, and whose labels are a consensus between the two original ontologies, e.g. "(grass; herbaceous)". This consensus between the two ontologies is minimal, in that it is much more efficient and meaningful than the mere cross-product of the two ontologies.

Therefore, ontologies are not made solely for Internet applications or e-business, but also for addressing real world queries: "*What there is*", or "*The nature of being*", as discussed by the philosopher W. O. Quine and others with an interest in ontology as a branch of philosophy which deals with the science of what is, what exists, and associated concerns. These fundamental questions have confronted geo-information scientists since the beginning of their work.

Figure 6. Consensus graph for the fusion of the two ontologies.

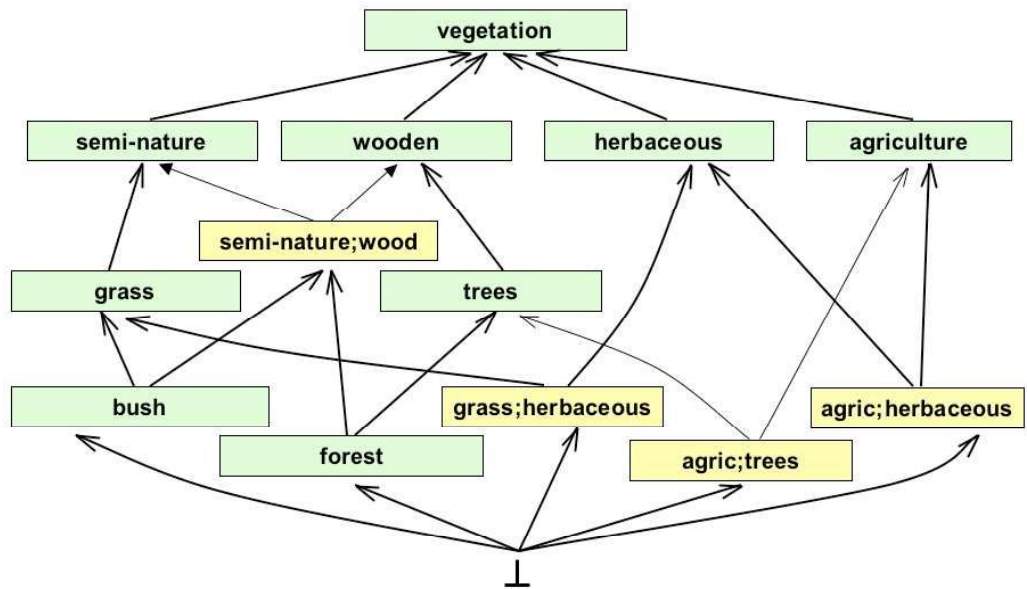

**H. Conclusion.**

Some 30 years ago, the conjunction of the multiplication of remote sensing imagery (Landsat 1976, then SPOT), the early stages of desktop image processing, the automation of the cartographic processing, and a rapidly increasing computing power – *doubling every two years* according to Moore's law, version 2 – offered an opportunity to collect, merge, and process an enormous amount of data, an amount larger than that ever collected and processed by machines.

The engineers and researchers involved in geo-information were, consequently, on the leading edge for the development of tools and methods that eventually are part of what today is termed big data.

Finally, in closing, a tribute.

It is appropriate to recognize the expertise that geo-information specialists have developed in data engineering. Decision-makers are increasingly relying on numbers to prepare and make their decisions. However, it appears that few of them are aware of the way these numbers are produced, that is, data are the result of many small decision-making processes at all stages. Data are not Facts, but Acts.

From the signal to the semantics of information, the science of geo-information has confronted, and been confronted by, a very large range of issues, and has brought its contribution to many data models, and many algorithms.

Dealing on an everyday basis with data and models, but dealing basically with the real world, the geo-information scientist must continue to seek out the right tools and representations, and thereby continue to pioneer advances in data engineering.

**I. References.**

Allen, J.1983. Maintaining knowledge about temporal intervals. In *Communications of the ACM*. 26 November: 832–843.

Cavarroc, M.-A., B. Salem, and R. Jeansoulin. 2004. Modeling landuse changes using Bayesian networks. *22nd IASTED Intl. Conf. Artificial Intelligence and Applications*, Innsbruck, Austria.

Cortes, C., and V. Vapnik. 1995. Support-vector networks. *Machine Learning*, 20, 1995. http://www.springerlink.com/content/k238jx04hm87j80g/

Degenne, P., D. Lo Seen, D. Parigot, R. Forax, A.Tran, L. Ait, A. Lahcen, O.Curé, and R. Jeansoulin. 2009. Design of a domain specific language for modelling processes. *Ecological Modelling*, 220(24): 3527-3535.

Devillers, R. and R. Jeansoulin. 2006. *Fundamentals of Spatial Data Quality*. ISTE Publishing Company, London UK.

Diday, E. 1973. The dynamic clusters method in nonhierarchical clustering. *International Journal of Computer & Information Sciences*, March 1973, Volume 2, Issue 1, pp 61-88.

Edwards, G., and R. Jeansoulin. 2004. Data fusion: From a logic perspective with a view to implementation. *International Journal of Geographical Information Science*, 18(4):303-307.

Euzenat, J., B. Christian, R. Jeansoulin, J. Revault, and S. Schwer, 1997. Raisonnement spatial et temporel. *Bulletin de l'Association Française pour l'Intelligence Artificielle,* Vol.2: 2-13.

Ferrucci, D., *et al.* 2010. Building Watson: An overview of the DeepQA Project. *AI Magazine 31(3) 59-79.* (see: http://www.aaai.org/Magazine/Watson/watson.php)

Goodchild, M. 2010. Twenty years of progress: GIScience in 2010. *Journal of Spatial Information Science* (1); 3–20.

Gruber, T and G. Olsen. 1994. An ontology for engineering mathematics. *Proceedings of the Fourth International Conference on Principles of Knowledge Representation and Reasoning*: 258–269.

Halevy, A. 2005. Why your data don't mix? *ACM Queue 3(8)*. (retrieved at: http://homes.cs.washington.edu/~alon/files/acmq.pdf )

Jeansoulin, R., O. Papini, H. Prade, and S. Schockaert, 2010. *Methods for Handling Imperfect Spatial Information*. Springer Berlin Heidelberg.

Jeansoulin, R. and N. Wilson. 2002. Quality of geographic information: Ontological approach and artificial intelligence tools in the Revigis project. *EC-GI& GIS Workshop*. 12.

Jeansoulin, R., Y. Fontaine, and W. Frei.1981. Multitemporal segmentation by means of fuzzy sets. *7th LARS Symposium on Machine Processing of Remotely Sensed Data, with Special Emphasis on Range, Forest, and Wetlands Assessment*. Purdue University: 336-340.

Moore, E.G. 1975. *Progress in Digital Integrated Electronics.* IEEE.

Olioso, A., *et al*. 1998. Spatial aspects in the Alpilles-ReSeDA project. *Scaling and Modeling in Forestry: Application in Remote Sensing and GIS.* D Marceau, ed. Université de Montréal: 92–102.

Pham, T.T., V Phan-Luong, and R. Jeansoulin, 2004. Data quality based fusion: Application to the land cover. *7th International Conference on Information Fusion* (FUSION '04).